\documentclass[a4paper,11pt]{article}
\usepackage{jinstpub} % for details on the use of the package, please see the JINST-author-manual
\usepackage{lineno}
\usepackage{amsmath}
\usepackage{bm}
\usepackage{braket}
\usepackage{caption}
\usepackage{subcaption}
\title{\boldmath Hanle Effect for Lifetime Analysis: Li-like Ions}

\author[a,b,1]{Jan Richter\note{Corresponding author.},}
\author[c,d,e]{Moto Togawa,}
\author[c]{Jos\'e R.~{Crespo L\'opez-Urrutia}}
\author[a,f]{and Andrey Surzhykov}%
\affiliation[a]{Physikalisch–Technische Bundesanstalt,\\ D–38116 Braunschweig, Germany}%
\affiliation[b]{Institut f\"ur Theoretische Physik, Leibniz Universität Hannover,\\ Appelstraße 2, 30167 Hannover, Germany}
\affiliation[c]{Max-Planck-Institut f\"ur Kernphysik,\\ Saupfercheckweg 1, 69117 Heidelberg, Germany}%
\affiliation[d]{European XFEL,\\ Holzkoppel 4, 22869 Schenefeld, Germany}%
\affiliation[e]{Heidelberg Graduate School for Physics, Ruprecht-Karls-Universität Heidelberg,\\ Im Neuenheimer Feld 226, 69120 Heidelberg, Germany}%
\affiliation[f]{Institut f\"ur Mathematische Physik, Technische Universität Braunschweig,\\ Mendelssohnstrasse 3, D-38106 Braunschweig, Germany}

\emailAdd{jan.richter@ptb.de}

\abstract{
Accurate lifetime measurements of excited states of highly charged ions (HCIs) are essential for advancing diagnostics in both laboratory and astrophysical plasmas, especially in the X-ray regime. The Hanle effect, which utilizes external magnetic fields to modify photon scattering patterns, provides a powerful technique for these measurements. Previously, this method has been  successfully employed for He-like ions. 
Here, we present a theoretical study of the prospects of the Hanle effect for lifetime determinations of Li-like ions. Our results highlight the potential for plasma diagnostics and X-ray spectral analysis.
}


\begin{document}
\maketitle
\flushbottom

\section{Introduction}
\label{sec: Introduction}

%The measurement of lifetimes in high-temperature plasmas is crucial for advancing the understanding of both laboratory-generated and astrophysical plasma environments. 
Highly charged ions, as for instance He-like and Li-like ions, play a crucial role in the study of both astrophysical and laboratory plasmas~\cite{ChandraXMMNewton,Yamamoto2010,Hirsch2019}. 
Therefore, accurate experimental and theoretical data for such ions, including oscillator strengths and lifetimes, are essential for plasma diagnostics.
In a recent study, the Hanle effect was employed to measure lifetimes of excited $1\mathrm{s} n\mathrm{p}$ $^1\mathrm{P}_1$ states of He-like nitrogen ions, providing valuable data on soft X-ray transitions and addressing gaps in knowledge regarding fast radiative processes in these ions~\cite{Togawa2024}.
The theoretical framework of these measurements is rooted in resonant elastic photon scattering, a fundamental process of atom-light interaction, which has been extensively discussed in the literature~\cite{Serbo22,Volotka22,Samoilenko2020}. In particular, an external magnetic field, like that present in electron beam ion traps (EBITs), alters the angular distribution of the emitted photons due to the Hanle effect~\cite{Togawa2024,hanle1924,Avan1975,moruzzi2013hanle}. 
The lifetime of excited states can be extracted with high precision from observations of the resulting angular distribution of scattered light.\\
The resonant scattering in the above mentioned experiment took place via the transition from the $1\mathrm{s}^2$ $^1\mathrm{S}_0$ ground state to one of the $1\mathrm{s} n\mathrm{p}$ $^1\mathrm{P}_1$ excited states.
The present work extends the theoretical investigation of lifetime measurements based on the Hanle effect to the scattering of photons by ions in the ground state with angular momentum $J_i=1/2$ via excited states with angular momentum $J_\nu=1/2$ or $J_\nu=3/2$, as it can be observed, for instance, in Li-like ions.
Our analysis starts in Sec.~\ref{sec: Geometry} with establishing the appropriate scattering geometry, based on the experimental setup described in Ref.~\cite{Togawa2024}.
The theoretical description of the scattering process, which is relying on a second-order perturbation approach, is given then in Sec.~\ref{sec: Resonant Scattering Theory}.
Here, in particular, we discuss the angle-differential cross section for the scattering of linearly polarized incident light.
In Sec.~\ref{sec: Results}, we show that the ratio of cross sections for outgoing photons emitted either in parallel or perpendicular direction to the polarization vector of incident ones is very sensitive both to the external magnetic field and the lifetime of an intermediate $J_\nu=3/2$ state. Lifetime measurements of the excited states of Li-like ions can be performed by employing this sensitivity for the range $10^{-13}$\,s $\leq \tau_\nu \leq 10^{-10}$\,s. A summary of these results and their potential for future experimental applications are discussed in Sec.~\ref{sec: Conclusion}.
Relativistic units ($\hbar = c = m_e = 1$) are used throughout this paper unless stated otherwise.

\section{Geometry}
\label{sec: Geometry}
\begin{figure}
    \centering
    \begin{subfigure}{0.49\textwidth}
    \centering
    \includegraphics[width=1.1\linewidth]{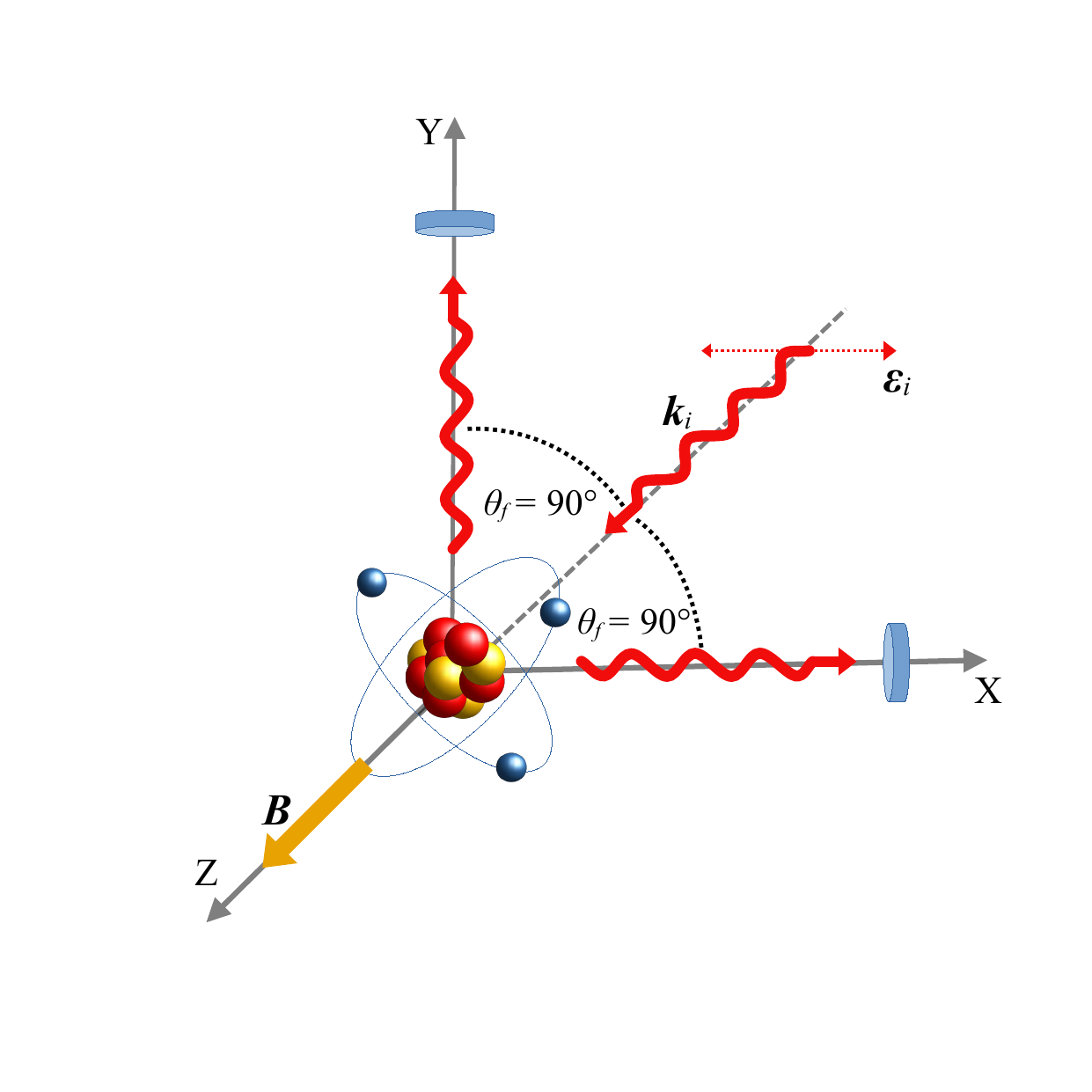}
    \caption{}
    \label{Fig: Geometry}
    \end{subfigure}
    \begin{subfigure}{0.49\textwidth}
    \centering
    \includegraphics[width=1.1\linewidth]{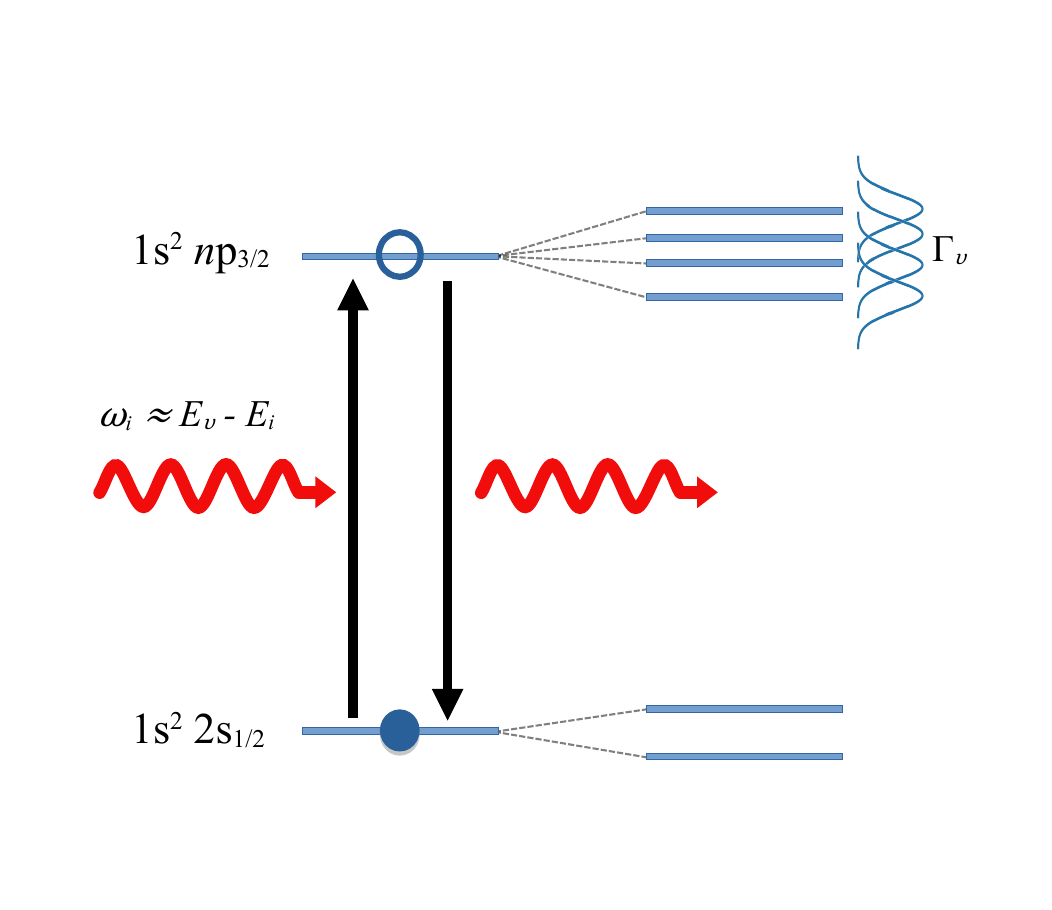}
    \caption{}
    \label{Fig: Levelscheme}
    \end{subfigure}
    \caption{(a) The geometry of the scattering process with the incident radiation propagating parallel to the external magnetic field in Z-direction. The polarization vector of the incoming photons is oriented along the X-axis while the scattered photons are detected in X- and Y-direction. (b) A schematic illustration of the resonant scattering via the $1\mathrm{s}^2 2\mathrm{s}$ $^2\mathrm{S}_{1/2} \to 1\mathrm{s}^2 n\mathrm{p}$ $^2\mathrm{P}_{3/2}$ transition in Li-like ions.}
    \label{Fig: Geometry and Levelscheme}
    \label{fig:enter-label}
\end{figure}

Before analyzing the resonant scattering process, it is essential to establish first its geometry, which is based on the experimental setup described in Ref.~\cite{Togawa2024}.
In this setup, displayed in Fig.~\ref{Fig: Geometry}, the magnetic field $\bm B$ is aligned with the propagation direction of the incoming photon beam, defining the quantization Z-axis. The incoming radiation is linearly polarized in the X-Z-plane, and the emitted photons are detected at a polar angle $\theta_f = 90^\circ$. Detection occurs either within the polarization plane of the incoming radiation ($\phi_f=0$) or perpendicular to it ($\phi_f=90^\circ$).

\section{Resonant Scattering Theory}
\label{sec: Resonant Scattering Theory}

In the low-intensity regime, the interaction between light and an ion can be described within the perturbation theory. For the photon scattering, this yields the well-known second-order amplitude, extensively discussed in the literature~\cite{kane1986,Strnat2021,Strnat2022,Manakov_2000,Serbo22}. When the initial photon energy $\omega_i$ closely matches the energy of the transition from the ground state to some intermediate ionic state, $\omega_i\approx E_\nu-E_i$, the resonant approximation~\cite{Serbo22,Volotka22} simplifies the scattering amplitude to:
\begin{equation}
\label{Eq.: Scattering matrix element resonant}
\mathcal{M}^{\text{res}}_{M_f,M_i} = \alpha \sum_{M_\nu} \frac{\braket{f|\hat{\mathcal{R}}^\dagger(\bm{k}_f,\bm{\epsilon}_f)|\nu} \braket{\nu|\hat{\mathcal{R}}(\bm{k}_i,\bm{\epsilon}_i)|i}}{E_i - E_\nu + \omega_i + i \Gamma_\nu / 2}.
\end{equation}
Here, $\ket{i} = \ket{\gamma_i J_i M_i}$, $\ket{\nu} = \ket{\gamma_\nu J_\nu M_\nu}$, and $\ket{f} = \ket{\gamma_f J_f M_f}$ represent the initial, intermediate, and final states, respectively, with total angular momentum $J$, its projection $M$ and $\gamma$ denoting all remaining quantum numbers for a unique characterization of the states. Moreover, $\alpha$ is the fine structure constant, $\Gamma_\nu$ is the natural width of the intermediate state, and the operators $\hat{\mathcal{R}}$ and $\hat{\mathcal{R}}^\dagger$ describe photon absorption and emission, respectively.
In the Coulomb gauge, the absorption operator can be written as:
\begin{equation}
\hat{\mathcal{R}}(\bm{k}, \bm{\epsilon}) = \sum_q \bm{\alpha}_q \cdot \bm{\epsilon} e^{i \bm{k} \bm{r}_q},
\end{equation}
where $\bm{\alpha}$ denotes the vector of Dirac matrices, and $\bm{\epsilon}$ and $\bm{k}$ represent the photon's polarization and momentum vectors, respectively. Matrix elements of these operators are typically evaluated with the help of a multipole expansion and by using the Wigner-Eckart theorem~\cite{Serbo22,Volotka22,Samoilenko2020}.\\
Similar as in Refs. \cite{Togawa2024, Stenflo1998}, we can use the amplitude~\eqref{Eq.: Scattering matrix element resonant} to describe resonant photon scattering in the presence of an external magnetic field. 
In order to achieve this, one has to modify the energies of initial and intermediate states in the denominator of the scattering amplitude~\eqref{Eq.: Scattering matrix element resonant} as: 
\begin{equation}
\label{Eq: energies modified}
\begin{aligned}
    E_i &= E_i^{(0)} + \Delta E_{Z,i} =E_i^{(0)}+ M_i g_i \mu_B B, \\
    E_\nu &= E_\nu^{(0)} + \Delta E_{Z,\nu}  =E_\nu^{(0)}+ M_\nu g_\nu \mu_B B.
\end{aligned}
\end{equation}
Here, $E^{(0)}$ denotes the unperturbed energies and $\Delta E_Z$ is the additional Zeeman shift, where $g$ is the Landé $g$-factor, $\mu_B$ is the Bohr magneton, and the magnetic field strength is given by $B=\left|\bm B\right|$.
By employing the matrix element~\eqref{Eq.: Scattering matrix element resonant} alongside the modified energies from Eq.~\eqref{Eq: energies modified}, one derives the differential cross section for resonant scattering
\begin{equation}
\label{Eq: differential cross section}
     \frac{\mathrm{d}\sigma}{\mathrm{d}\Omega}\left(\theta_f,\phi_f,\omega_i\right) = \frac{1}{2 J_i +1} \sum_{M_i ,M_f, \bm{\epsilon}_f}\left|\mathcal{M}^{res}_{M_f,M_i}\right|^2.
 \end{equation}
This cross section depends on the angles $\left(\theta_f,\phi_f\right)$, defined relative to the wave vector $\bm{k}_i$ of the initial radiation and its polarization $\bm{\epsilon}_i$, see Fig.~\ref{Fig: Geometry}. Moreover, it is assumed that the ion's initial state is unpolarized and that there is no observation of the magnetic sublevels $\ket{\gamma_f J_f M_f}$ or the polarization of scattered light.

\section{Results}
\label{sec: Results}
As follows from Eqs.~\eqref{Eq.: Scattering matrix element resonant}-\eqref{Eq: differential cross section}, the angle-differential cross section of resonant photon scattering is sensitive to the strength of an external magnetic field.
In Ref.~\cite{Togawa2024}, this $B$-field sensitivity has been employed to study lifetimes of excited $1\mathrm{s} n\mathrm{p}$ $^1\mathrm{P}_1$ states of He-like N ions. 
In the present work, we will extend this study towards $J_i=1/2 \to J_\nu=1/2$ and $J_i=1/2 \to J_\nu=3/2$ transitions. We focus in particular on the cross sections either parallel or perpendicular to the polarization vector of incident light:
\begin{equation}
    \begin{aligned}
        \sigma_\perp\left(\omega_i\right) &\equiv \frac{\mathrm{d}\sigma}{\mathrm{d}\Omega}\left(\theta_f=90^\circ,\phi_f=90^\circ,\omega_i\right), \\ 
        \sigma_\parallel\left(\omega_i\right) &\equiv \frac{\mathrm{d}\sigma}{\mathrm{d}\Omega}\left(\theta_f=90^\circ,\phi_f=0,\omega_i\right).
    \end{aligned}
\end{equation}
We will investigate how the ratio of these cross sections, $\sigma_\parallel/\sigma_\perp$, depends on the external magnetic field and lifetime of an excited ionic state. 
Moreover, we neglect any effects of hyperfine structure on the scattering process, by restricting our analysis to ions with nuclear spin $I=0$.

\subsection{Scattering via a $J_\nu=1/2$ state}

In this subsection, we briefly discuss a $J_i=1/2 \to J_\nu=1/2 \to J_i=1/2$ transition, such as for example the photon scattering by Li-like ions in the $1\mathrm{s}^2 2\mathrm{s}$ $^2\mathrm{S}_{1/2}$ ground state via an intermediate $1\mathrm{s}^2 n\mathrm{p}$ $^2\mathrm{P}_{1/2}$ level. Using Eqs.~\eqref{Eq.: Scattering matrix element resonant}-\eqref{Eq: differential cross section}, one finds that for $J_i=J_\nu=1/2$, the angular distribution of scattered photons in the field-free case, $B=0$, is isotropic~\cite{Volotka22}, resulting in the ratio $ \sigma_\parallel/\sigma_\perp = 1$. 
This isotropy persists even in the presence of an external magnetic field when accounting for the modified energies from Eq.~\eqref{Eq: energies modified}. Consequently, the ratio remains $ \sigma_\parallel/\sigma_\perp = 1$, making it impossible to extract the lifetimes of excited states from the analysis of the angular distribution of emitted photons.

\noindent One should note that a different experimental setup was proposed in Refs.~\cite{zimmermann1975,Feichtner1967}, which allows lifetime determination utilizing the Hanle effect also for $J_i=1/2 \to J_\nu=1/2 \to J_i=1/2$ transitions.
This setup relies on the scattering of circularly polarized incoming light as well as on the detection of circularly polarized outgoing photons. Moreover, a scan over the external magnetic field strength is performed to observe the zero-field level crossing of the degenerate magnetic sublevels of the excited state.
This scenario, requires the production and detection of circularly polarized radiation, which is well established in the visible light domain, but is more challenging in the EUV and X-ray regime.

\subsection{Scattering via a $J_\nu=3/2$ state}

\begin{figure}
    \centering
    \includegraphics[width=0.9\linewidth]{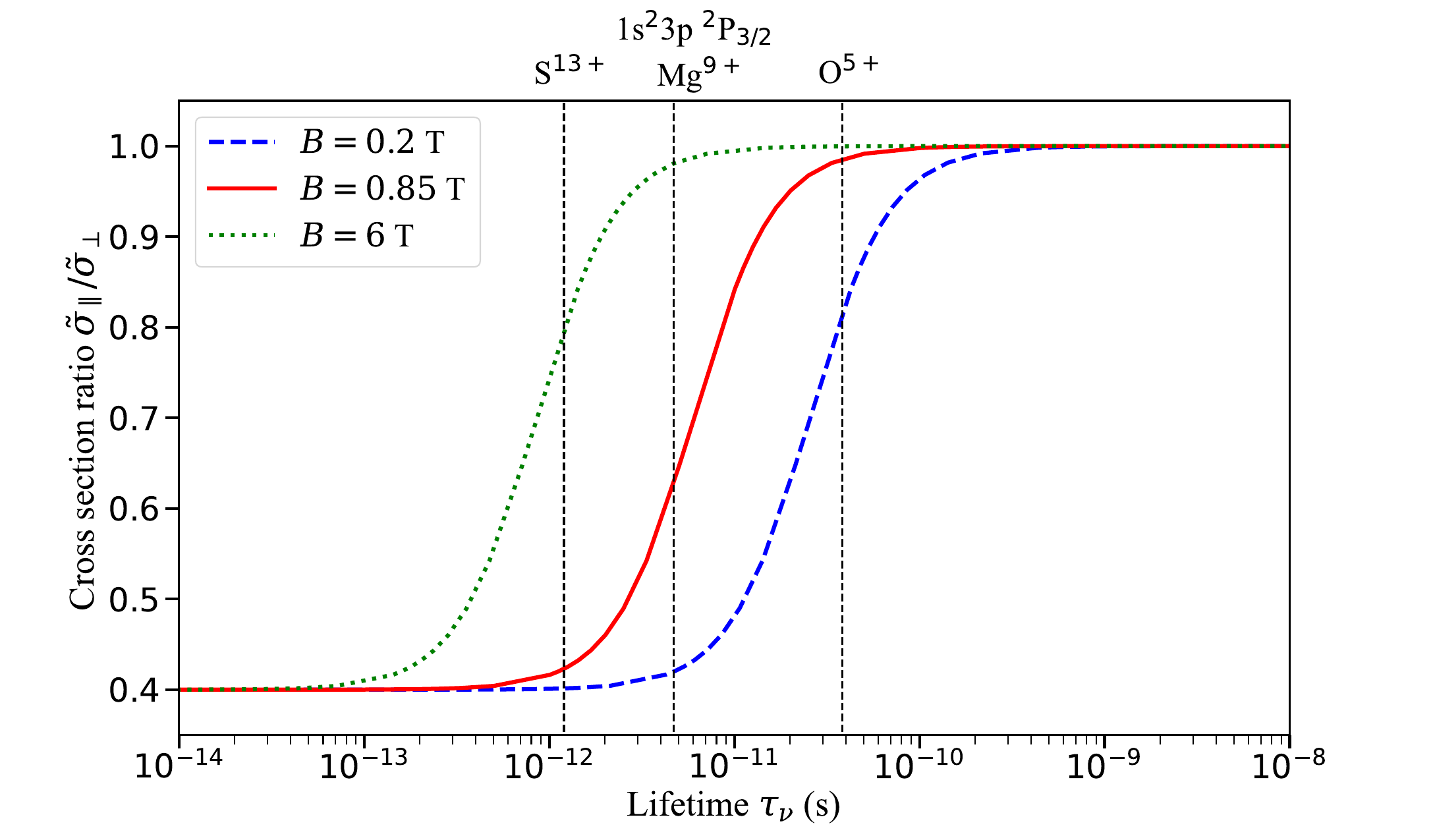}
    \caption{The ratio of frequency averaged cross sections for photons scattered within ($\tilde\sigma_\parallel$) or perpendicular ($\tilde\sigma_\perp$) to the polarization plane of the incident radiation as a function of the lifetime of the excited state. The calculations were performed for external magnetic field strengths of $B=0.2$\, T (blue dashed line), $B=0.85$\,T (red solid line), and $B=6$\,T (green dotted line). The vertical line present the lifetimes of the $1\mathrm{s}^2 3\mathrm{p}$ $^2\mathrm{P}_{3/2}$ state in Li-like O, Mg and S ions. }
    \label{fig: results}
\end{figure}

A more promising scenario involves $J_i=1/2 \to J_\nu=3/2 \to J_i=1/2$ transitions, such as the resonant scattering via  excited $1\mathrm{s}^2 n\mathrm{p}$ $^2\mathrm{P}_{3/2}$ states in Li-like ions. For the field-free case, $B=0$, the angle-differential cross section of emitted photons for linearly polarized initial radiation is given by~\cite{Volotka22}:
\begin{equation}
    \frac{\mathrm{d}\sigma}{\mathrm{d}\Omega}\left(\theta_f,\phi_f,\omega_i\right)=\sigma_0\left(\omega_i\right)\left(17+3 \cos\left(2\theta_f\right)-6 \sin^2\left(\theta_f\right) \cos\left(2 \phi_f\right)\right),
\end{equation}
thus leading to the ratio $\sigma_\parallel/\sigma_\perp = 2/5$. When a magnetic field is applied, the cross section $\mathrm{d}\sigma/\mathrm{d}\Omega$ and, hence, the ratio $\sigma_\parallel/\sigma_\perp$ is modified. While the exact analytical expression for the cross section ratio in the presence of an external magnetic field is cumbersome, and therefore is not presented here, we investigate it numerically.
To perform this quantitative analysis, we focus below on the resonant $1\mathrm{s}^2 2\mathrm{s}$ $^2\mathrm{S}_{1/2} \to 1\mathrm{s}^2 n\mathrm{p}$ $^2\mathrm{P}_{3/2}$ scattering. 
As seen from Eq.~\eqref{Eq: energies modified}, the computation of the Zeeman splitting, and hence the $g$-factors, of the involved states is needed.
For low-$Z$ ions, the $g$-factors can be approximated as $g_{^2\mathrm{S}_{1/2}}=2$ and $g_{^2\mathrm{P}_{3/2}}=4/3$. 
Moreover, to account for a realistic experimental scenario of a finite frequency width of the incoming radiation, we average the cross section~\eqref{Eq: differential cross section} over a Gaussian distribution:
\begin{equation}
\label{eq: cross section freq averaged}
    \tilde \sigma_{\perp,\parallel} = \int \sigma_{\perp,\parallel}\left(\Delta_i\right)\frac{1}{\Gamma_\omega \sqrt{2\pi}} e^{-\frac{\Delta_i^2}{2\Gamma_\omega^2}} \, \mathrm{d}\Delta_i,
\end{equation}
where $\Delta_i=\omega_i-(E_\nu^{(0)}-E_i^{(0)})$ is the detuning and $\Gamma_\omega = 0.1$\,eV is the width of the incident light.\\
The ratio $\tilde\sigma_\parallel/\tilde\sigma_\perp$ of the frequency averaged cross sections depends both on the strength of the external magnetic field and the natural width $\Gamma_\nu$ of an excited state, directly related to its lifetime $\tau_\nu=\hbar/\Gamma_\nu$.
In order to better understand this sensitivity, we display in Fig.~\ref{fig: results} the ratio $\tilde\sigma_\parallel/\tilde\sigma_\perp$ calculated for different magnetic field strengths, $B=0.2$\,T (blue dashed line), $B=0.85$\,T (red solid line), $B=6$\,T (green dotted line), and as a function of the lifetime $\tau_\nu$. Hence, we formally use $\tau_\nu$ as a free parameter of our calculations with the goal to specify a feasible range for lifetime measurements. For reference, by the vertical lines in Fig.~\ref{fig: results} we mark as examples actual lifetimes of $1\mathrm{s}^2 3\mathrm{p}$ $^2\mathrm{P}_{3/2}$ states in Li-like O, Mg, and S ions.\\
As seen from Fig.~\ref{fig: results}, the cross section ratio $\tilde\sigma_\parallel/\tilde\sigma_\perp$  is constant for small and large lifetimes $\tau_\nu$, but exhibits a steep growth in the intermediate region $10^{-13}$\,s $\leq \tau_\nu \leq 10^{-10}$\,s.
To understand this behaviour, we recall that for small lifetimes the Zeeman splitting in Eqs.~\eqref{Eq.: Scattering matrix element resonant} and \eqref{Eq: energies modified} becomes negligible compared to the natural width, $\Delta E_Z \ll \Gamma_\nu=\hbar/\tau_\nu$, thus recovering the field-free result $\tilde\sigma_\parallel/\tilde\sigma_\perp=2/5$.
On the other side, larger lifetimes correspond to reduced widths $\Gamma_\nu$, which are much smaller than the Zeeman splitting. In this regime, the magnetic sublevels $\ket{^2\mathrm{P}_{3/2} M_\nu}$ are well separated and do not interfere, thus leading to insensitivity to variations of $B$ and $\tau_\nu$.\\
As mentioned already above, the cross section ratio is very sensitive to the lifetime of the excited state in the intermediate lifetime regime $10^{-13}$\,s $\leq \tau_\nu \leq 10^{-10}$\,s.
This strong sensitivity is a clear manifestation of the Hanle effect, where a partial overlap of Zeeman sublevels can significantly modify angular distribution and polarization of scattered photons~\cite{Togawa2024,hanle1924,Avan1975,moruzzi2013hanle}. Consequently, this regime is the most attractive for lifetime measurements. Moreover, as seen from Fig. \ref{fig: results}, the accessible range of lifetimes can be further fine-tuned by a variation of the magnetic field strength.
For example, the lifetimes in the range of $10^{-12}$\,s $\leq \tau_\nu \leq 3\times 10^{-11}$\,s are within reach for $B=0.85$\,T, which can be realized with the PolarX-EBIT~\cite{Togawa2024,Micke2018}.
This range corresponds, for example, to the lifetimes of the $1\mathrm{s}^2 3\mathrm{p}$ $^2\mathrm{P}_{3/2}$ state of Li-like ions with $9\leq Z\leq 16$. In order to investigate lifetimes of heavier ions one would need to apply a stronger magnetic field of $B=6$\,T, which is available with FLASH-EBIT~\cite{FLASH}. On the other side, to access the lifetime range of lighter elements one can apply a smaller $B$ as shown by the blue dashed line in Fig~\ref{fig: results}. \\  
So far our analysis of the $^2\mathrm{S}_{1/2} \to$ $^2\mathrm{P}_{3/2} $ scattering has been restricted to the electric dipole approximation.
However, this scattering process may proceed also via a magnetic quadrupole channel. As shown in Ref.~\cite{Volotka22}, this channel alters the angular and polarization properties of scattered light by up to $15\%$ for high-Z ions but is negligible for low- and medium-Z ions. 
A complete analysis of the magnetic quadrupole contribution’s impact will be addressed in future work.

\section{Conclusion}
\label{sec: Conclusion}
We presented a theoretical investigation of the resonant scattering of linearly polarized radiation by highly charged ions. Special attention was paid to the impact of an external magnetic field on the angle-differential cross section. 
It was demonstrated in our previous study~\cite{Togawa2024} that this cross section is very sensitive not only to the magnetic field strength but also to the lifetime of the intermediate (excited) state of the scattering process.
In this work, we extended this method, based on the well known Hanle effect, to the case of $J_i=1/2 \to J_\nu=3/2 \to J_i=1/2$ scattering, which may occur, for example, in Li-like ions.
Based on our theoretical analysis and numerical calculations, we argue that angular resolved measurements of scattered radiation can be utilized to determine lifetimes of excited states of Li-like ions.
It is shown that, for realistic experimental parameters, the lifetimes in the range $10^{-13}$\,s $\leq \tau_\nu \leq10^{-10}$\,s can be studied.
The proposed method holds promise for applications in plasma diagnostics and the analysis of astrophysical spectra, particularly where lithium-like ions play a significant role. With further refinement, this technique could provide vital lifetime data, enhancing the interpretive power of observations from facilities such as XRISM, XMM-Newton and Chandra.

\acknowledgments

J.R. and A.S. acknowledge funding by the Deutsche Forschungsgemeinschaft (DFG, German Research Foundation) under Germany’s Excellence Strategy – EXC-2123 QuantumFrontiers – 390837967.

\bibliographystyle{JHEP}
\bibliography{main.bib}
\end{document}